\documentclass[a4paper,11pt]{article}
\usepackage{jcappub}
\newcommand{\figurewidth}{0.8\textwidth}

\abstract{
 An epoch of matter domination in the early universe can enhance the primordial stochastic gravitational wave signal, potentially making it detectable to upcoming gravitational wave experiments.  However, the resulting gravitational wave signal is quite sensitive to the end of the early matter-dominated epoch.  If matter domination ends gradually, a cancellation results in an extremely suppressed signal, while in the limit of an instantaneous transition, there is a resonant-like enhancement.  The end of the matter dominated epoch cannot be instantaneous, however, and previous analyses have used a Gaussian smoothing technique to account for this, and consider only a limited regime around the fast transition limit.  In this work, we present a study of the enhanced gravitational wave signal from early matter domination without making either approximation and show how the signal smoothly evolves from the strongly suppressed to strongly enhanced regimes.}

\begin{document}

\title{
Gravitational Wave Signals From Early Matter Domination: Interpolating Between Fast and Slow Transitions
}

\author[a]{Matthew Pearce}
\affiliation[a]{School of Physics and Astronomy, Monash University, Melbourne 3800 Victoria, Australia}
\emailAdd{matthew.pearce1@monash.edu}

\author[b]{Lauren Pearce}
\affiliation[b]{Pennsylvania State University-New Kensington, New Kensington, PA 15068, USA}
\emailAdd{lpearce@psu.edu}

\author[c]{Graham White}
\affiliation[c]{School of Physics and Astronomy, University of Southampton, Southampton SO17 1BJ, United Kingdom}
\emailAdd{g.a.white@soton.ac.uk}

\author[a]{Csaba Bal\'azs}
\emailAdd{csaba.balazs@monash.edu}

\date{\today}

\maketitle

\section{Introduction}

The recent detection of gravitational wave (GW) signals has opened a new epoch in observational astronomy.  While current observations have mostly expanded our knowledge of black hole and neutron star populations~\cite{LIGOScientific:2014qfs,LIGOScientific:2016aoc,LIGOScientific:2017vwq,LIGOScientific:2017ync}, future experiments are expected to also probe early universe cosmology~\cite{Kosowsky:1992rz,Caprini:2019egz,Vilenkin:1981zs,Dunsky:2021tih}, including early matter-dominated (MD) epochs~\cite{Assadullahi:2009nf,Alabidi:2013lya,Kohri:2018awv,Inomata:sudden,Inomata:gradual}.

Primordial curvature perturbations, such as those produced by inflation, produce GWs at second order in perturbation theory~\cite{Kohri:2018awv}.  The signal strength of these perturbations depends on the cosmological history of the universe, particularly through the gravitational potential.  During an MD epoch, the gravitational potential does not decay, even on subhorizon scales, in contrast to a radiation-dominated (RD) epoch.  Consequently, the metric perturbations sourced at second order by the primordial curvature perturbations are enhanced as compared to the RD scenario.

Such an early MD epoch can occur if the energy density of the universe is dominated by either heavy quasi-stable objects or a rapidly oscillating, but decaying, scalar field.  Heavy objects can be motivated in extensions of the Standard Model (for example, moduli fields~\cite{Coughlan:1983ci,deCarlos:1993wie,Banks:1993en,Banks:1995dt}), as part of a dark sector (for example,~\cite{Pospelov:2007mp,Arkani-Hamed:2008hhe,Berlin:2016vnh}), or as extended objects, including primordial black holes~\cite{Inomata:pbh,Papanikolaou:2022chm} and non-topological solitons~\cite{White:2021hwi,Kasuya:2022cko}.  Currently, there are no constraints on the existence or duration of such an epoch, provided that it ended before nucleosynthesis~\cite{Amin:2014eta}.

Since an early MD epoch can arise in such a varied array of models~\cite{Harigaya:2023mhl,Datta:2023xpr,Borah:2023iqo,Gehrman:2023esa,Domenech:2023mqk,Kasuya:2022cko,Bhaumik:2022zdd,Chen:2022dah,ReyIdler:2022unr,Bhaumik:2022pil,Papanikolaou:2022chm,Kozaczuk:2021wcl,White:2021hwi,Haque:2021dha,Domenech:2021wkk,Dalianis:2020gup,Papanikolaou:2020qtd,Domenech:2020ssp,Bhaumik:2020dor,Inomata:pbh}
(for a review, see ref.~\cite{Domenech:2021ztg}), detailed studies of the associated GW signal are necessary.  In particular, scenarios may differ in how rapidly the early MD epoch ends.  Two limits have been well-studied in the literature, that of a slow transition (as compared to the Hubble time)~\cite{Inomata:gradual} and that of an instantaneous transition~\cite{Inomata:sudden}.  The resulting GW signals are quite different: in the fast transition limit, there is a resonant-like enhancement (known at the poltergeist mechanism~\cite{Inomata:pbh}); not only is this absent in the slow transition, but it is in fact replaced with a suppression.  This naturally leads to the question of how to interpolate between these two regimes.  This has been explored in the literature by smoothing the resonant enhancement with a Gaussian distribution of primordial black holes or Q-balls~\cite{Inomata:pbh,Kasuya:2022cko}, but a preferable approach would be to fully account for the finite duration of the reheating transition.  We present such an analysis in this work.

Our paper is arranged as follows.  In section~\ref{sec:modelling} we outline our approach, beginning with a brief review of the formalism for calculating the induced GWs.  We continue by outlining how we achieve an end to the early MD era that interpolates between the gradual and sudden regimes using a time dependent decay rate.  We conclude the section by explaining our treatment of the scale factor, Green's function and perturbations, highlighting the need for numerical calculations.  In section~\ref{sec:results} we present our calculation of the GW spectrum for varying transition duration, highlighting the disappearance of the resonant-like peak as the transition becomes more gradual as well as demonstrating the unphysical nature of oscillations in the spectrum seen in previous analysis of the gradual case~\cite{Inomata:gradual}.  In section~\ref{sec:conclusions} we present our conclusions.  We also give some technical details of the numerical methods we used to achieve accurate results in appendix~\ref{app:code}.

\section{Modelling end of early matter domination}
\label{sec:modelling}
In this section we outline the procedure for computing the power spectrum of GWs induced from curvature perturbations.  We then explain how we model a period of early matter domination that ends in a variable time frame via a parameterised, time dependent decay rate.

We work in the conformal Newtonian gauge with the perturbed metric
\begin{equation}
\label{eq:metric}
    ds^2 = a^2\left[-(1+2\Phi)d\eta^2+\left((1-2\Psi)\delta_{ij}+\frac{h_{ij}}{2}\right)dx^idx^j\right]
\end{equation}
where $a$ is the scale factor of the background FLRW metric, the conformal time $\eta = \int dt/a(t)$, $\Phi$ and $\Psi$ are the first order scalar perturbations and $h_{ij}$ are the tensor perturbations induced at second order.  We neglect anisotropic stress, which implies $\Psi=\Phi$, which we refer to as the gravitational potential.  Throughout the paper we will use primes to denote differentiation with respect to conformal time $\eta$.

Assuming a Gaussian distribution for the curvature perturbations, the power spectrum of GWs arising from scalar perturbations is given by~\cite{Kohri:2018awv,Inomata:2016rbd, Ananda:2006af,Baumann:2007zm}
\begin{equation}
\label{eq:spectra}
    \overline{\mathcal{P}_h(\eta,k)}=4\int_{0}^{\infty}dv\int_{|1-v|}^{1+v}du\left(\frac{4v^2-(1+v^2-u^2)^2}{4vu}\right)^2\overline{I^2(u,v,k,\eta)}\mathcal{P}_{\zeta}(uk)\mathcal{P}_{\zeta}(vk)
\end{equation}
where the overline denotes an oscillation average over time.  The power spectrum of curvature perturbations is given by
\begin{equation}
\label{eq:curvature}
{\mathcal{P}} _\zeta (k) = \Theta (k_{\rm max} - k) A_s \left( \frac{k}{k_\ast} \right) ^{n_s-1}   
\end{equation}
with $A_s=2.1\times 10^{-9}$ being the amplitude at the pivot scale, $n_s=0.97$ the spectral tilt, and $k_\ast=0.05 {\rm Mpc}^{-1}$ the pivot scale, where we take all of these values from ref.~\cite{Planck:2018vyg}.  We will also at times consider the scale invariant case with $n_s=1$.  The cutoff $k_{\rm max}$ should be taken to be the wavenumber of the mode that re-enters at the start of the MD era or the non-linear scale, whichever is smaller.  The application of linear perturbation theory is valid for $k_{\rm max}\leq 470/\eta_R$ \cite{Assadullahi:2009nf,Inomata:gradual} where $\eta_R$ is the time at which the universe transitions from the MD to RD era. (The specific definition of $\eta_R$ in the case of a gradual transition will be clarified below; we note that the time when the energy densities are equal may generally be different.)  To remain consistent with previous literature \cite{Inomata:gradual,Inomata:sudden} we conservatively take $k_{\rm max}=450/\eta_R$.  Further, the time dependence of the GWs is encapsulated in
\begin{equation}
\label{eq:I}
    I(u,v,k,\eta)=k\int_0^{\eta} d\Bar{\eta}\;\frac{a(\Bar{\eta})}{a(\eta)}kG_k(\eta,\Bar{\eta})f(u,v,k,\Bar{\eta})
\end{equation}
where the Green's function $G_k$ is the solution to the GW equation of motion
\begin{equation}
    \label{eq:greens}
    G''_k(\eta,\Bar{\eta})+\left(k^2-\frac{a''(\eta)}{a(\eta)}\right)G_k(\eta,\Bar{\eta})=\delta(\eta-\Bar{\eta}).
\end{equation}
The source function is given by the expression
\begin{align}
    f(u,v,k,\Bar{\eta})=\frac{3}{25(1+w)}&\Bigl(2(5+3w)\Phi(uk,\Bar{\eta})\Phi(vk,\Bar{\eta})\nonumber\\
    &+4\mathcal{H}^{-1}(\Phi'(uk,\Bar{\eta})\Phi(vk,\Bar{\eta})+\Phi(uk,\Bar{\eta})\Phi'(vk,\Bar{\eta}))\nonumber\\
    &+4\mathcal{H}^{-2}\Phi'(uk,\Bar{\eta})\Phi'(vk,\Bar{\eta})\Bigr)
    \label{eq:source}
\end{align}
where the equation of state is denoted by $w$, $\Phi$ is the transfer function of the gravitational potential, normalised so that $\Phi(x\rightarrow 0)=1$, and ${\cal H}=a^\prime/a$ is the conformal Hubble parameter.\footnote{We note that this convention is consistent with eq.~\eqref{eq:source}.  Other references, such as ref.~\cite{Kohri:2018awv}, normalise $\Phi$ differently but also correspondingly alter the $w$ dependence in the source (the only place where $\Phi$ enters the analysis).  Finally, we note that if the initial condition is set in the RD epoch that precedes the early MD epoch, then $\Phi_0 = 10 \slash 9$, so that $\Phi = 1$ at the start of the MD era; this is done in e.g., ref.~\cite{Inomata:pbh}.}  Finally, from the power spectrum, we can derive the GW abundance
\begin{align}
    \Omega_{\rm GW}(\eta,k)&=\frac{\rho_{\rm GW}(\eta,k)}{\rho_{\rm tot}(\eta)}\nonumber\\
    &=\frac{1}{24}\left(\frac{k}{\mathcal{H}(\eta)}\right)^2 \overline{\mathcal{P}_h(\eta,k)}.
\end{align}

Because previous analyses, whether of a rapid or slow transition, used piecewise approximations about $\eta_R$ for the scale factor, Green's function and gravitational potential~\cite{Inomata:gradual,Inomata:sudden,Inomata:pbh}, they decomposed the GW signal into three parts: one sourced from the RD era, one sourced from the MD era, and a cross term.  In the rapid scenario, the spectra is dominated by contributions sourced in the RD era, as the gravitational potential is matched at the reheating time with no decay.  In the case of a gradual transition, the contributions from the RD and MD eras are very similar, so the cancellation due to the cross term is severe.  We are interested in intermediate cases between the two extremes.  Although in our analysis we will avoid any piecewise approximations by numerically evolving the scale factor, Green's function and gravitational potential, we will still decompose our results into these components in Section \ref{sec:results} when we compare our results to previous work.

To model a reheating transition in the intermediate regime between gradual and rapid, we propose a time dependent decay rate with a parameter $\beta$ that controls the speed at which the matter evaporates.  Specifically we choose a decay rate of the form,
\begin{equation}
\label{eq:decay}
    \Gamma(\eta)=\Gamma_{max}(\tanh(\beta(\eta-\eta_{\Gamma}))+1)/2.
\end{equation}
While this decay rate is centred at the time $\eta_{\Gamma}$, the reheating transition will commence at time $\eta_R$ when the decay becomes efficient compared to the Hubble rate $\Gamma\sim\mathcal{H}$.  Consequently, to have a rapid decay with $\Gamma\gg \mathcal{H}$, it is necessary to have a time-evolving decay rate which rapidly increases.

In more detail, we note that $\eta_R$ depends not only on $\eta_{\Gamma}$ but also $\beta$ and $\Gamma_{\rm max}$.  For small enough $\beta$, the transition will occur during the rise of $\tanh$, in a region where the decay rate is not changing rapidly, resulting in a gradual transition.  Therefore, for fixed $\eta_{\Gamma}$, decreasing $\beta$ shifts the transition to earlier times.  The extreme limit of this is $\beta=0/\eta_R$ for which the decay rate becomes constant.  In the other limit, as $\beta\rightarrow\infty$, the decay rate approaches a step function which, given a large enough $\Gamma_{\rm max}$ such that $\Gamma(\eta_{\Gamma})\gg\mathcal{H}$, describes a rapid transition where all of the matter evaporates approximately instantaneously at $\eta_{\Gamma}$.

The matter energy density $\rho_m$, the radiation energy density $\rho_r$ and the scale factor $a$ will evolve according to the coupled Friedmann and continuity equations with our variable decay rate inserted as follows~\cite{Poulin:2016nat}
\begin{align}
\label{eq:fried_m}
    \rho_m'&=-(3\mathcal{H}+a\Gamma(\eta))\rho_m\\
\label{eq:fried_r}
    \rho_r'&=-4\mathcal{H}\rho_r+a\Gamma(\eta)\rho_m\\
\label{eq:fried_a}
    a'&=\sqrt{\frac{8\pi}{3M_{\rm Pl}^2}(\rho_m+\rho_r)}a^2
\end{align}
where $M_{\rm Pl}=1/\sqrt{G}$ is the Planck mass.
\begin{figure}[tbp]
    \centering
    \includegraphics[width=\figurewidth]{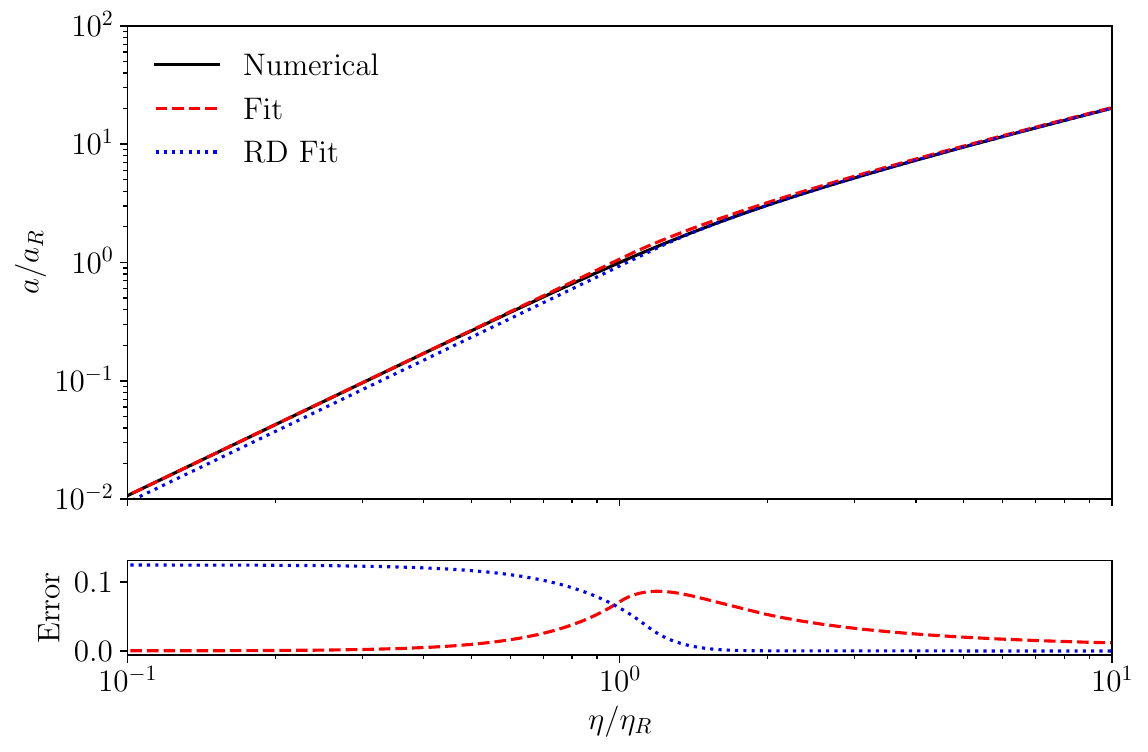}
    \caption{The time dependence of the scale factor in a gradual transition.  The numerical solution to eq.~\eqref{eq:fried_a} in black is contrasted to the instantaneous approximation eq.~\eqref{eq:scalefactor} in dashed red and dotted blue for the two choices of $\eta_R$ as discussed in the text.  The errors shown are the fractional errors in the each fit compared to the numerical computation, given by $\frac{|a-a_{\rm f}|}{a}$}
    \label{fig:ScaleFactor}
\end{figure}
In the solid black curve of figure~\ref{fig:ScaleFactor} we show the scale factor obtained from the Friedmann equations for a gradual transition with constant decay rate.  We see that it continuously evolves between the MD and RD regimes.  In contrast, we introduce an analytic fit $a_f$, valid for an exactly instantaneous transition
\begin{equation}
\label{eq:scalefactor}
    \frac{a_f(\eta)}{a_f(\eta_R)}=\begin{cases}
        \left(\tfrac{\eta}{\eta_R}\right)^2 & \eta<\eta_R \\
        2\tfrac{\eta}{\eta_R}-1 & \eta\geq\eta_R
    \end{cases}
\end{equation}
which is constructed so that $a_f$ and its first derivative are continuous at $\eta_R$.  Fits like this one have often been used in studies of GWs at second order, even in the case of gradual transitions (e.g. in ref.~\cite{Inomata:gradual,Inomata:sudden,Inomata:pbh}).  The fit has two free parameters $a_f(\eta_R)$ and $\eta_R$ which are to be determined by fitting onto two points of our numerical solution.  Note that $a_f(\eta_R)$ need not be the same as $a(\eta_R)$.  In fact enforcing this would make one of the points we fit to be in the middle of the transition which we find results in a poorer overall fit.

Since a definite transition time $\eta_R$ is a property of an instantaneous transition, when fitting to our smooth numerical calculation, different prescriptions for choosing $\eta_R$ can impact the error in the fit.  The two different methods we use are as follows.

In the red dashed curve of figure~\ref{fig:ScaleFactor} we compute $\eta_R$ and $a_f(\eta_R)$ by fitting to one point early in the MD era and another late in the RD era.  This captures the asymptotic behaviour in both eras, and as a result the calculation of $\eta_R$ is a good estimate for the time at which the universe transitions from being MD to RD.  Henceforth, when we refer to $\eta_R$ we use this method to compute it.  However, due to the gradual nature of the transition we see that this approach has as much as an eight percent error during the transition and the fit is poorer during the RD era.

Alternatively we perform the fit using two points in the RD era, where the fit is purely linear.  Specifically the first point is taken when $\rho_r=10^4\rho_m$ and the second much later, at the point where we stop integrating.  This is shown in the dotted blue curve of figure~\ref{fig:ScaleFactor}.  We see that the fit is very good during the RD era, with the actual size of the error being of order $10^{-7}$, while being more than 10\% during the MD era.  This fit will be useful for calculating an analytic expression for the perturbation $\Phi$ during the RD era.  From here on we use $\eta_{\rm RD}$ to indicate the value of $\eta_R$ in eq.~\eqref{eq:scalefactor} resulting from this approach.

We highlight that while we make some use of the fits, we do so in a way such that the errors introduced by their use are negligible.  The fit in the red dashed curved of figure~\ref{fig:ScaleFactor} is only used to estimate the time of the transition $\eta_R$.  This only enters the calculation when defining the non-linear cut-off scale $k_{\rm max}$ and when breaking the result down into contributions sourced from the MD era and those from the RD era.  The other fit is only used for the analytic expression for the gravitational potential late in the RD era after its relative error has dropped below $10^{-7}$.  During the MD era and the transition itself, when the error in the fits is largest, our calculation is fully numeric.  We note that this discussion has all been in the context of the gradual transition limit for which these issues are the most pronounced.  As the transition becomes more rapid, eq.~\eqref{eq:scalefactor} agrees better with the numerical solution.

Calculating the signal also requires the Green's function.  Equation~\eqref{eq:greens} can be solved analytically in the MD and RD regimes, and it is common in the literature to define a piecewise Green's function, changing between the two expressions at $\eta_R$.  This is done even in the analysis of gradual transitions, e.g.~ref.~\cite{Inomata:gradual}.  We improve on this by instead using a numerical solution for the Green's function which accurately captures the behaviour during the transition.  As we discuss below, we found that this eliminated unphysical oscillations (in frequency) in the resulting signal, which are present in the results of ref.~\cite{Inomata:gradual}.  In our approach, we numerically evolve eq.~\eqref{eq:greens} in $\eta$, from $\Bar{\eta}$, with initial conditions
\begin{equation}
    G_k(\eta\leq\Bar{\eta})=0,\;\; G_k'(\Bar{\eta},\Bar{\eta})=1  ,
\end{equation}
since we are looking for the retarded Green's function.  At sufficiently late times, we match onto the solution
\begin{equation}
\label{eq:G_decomposition}
    G_k(\eta_,\Bar{\eta})=E(\Bar{\eta})\cos{(k\eta)}+F(\Bar{\eta})\sin{(k\eta)}
\end{equation}
since in the RD limit $a''$ vanishes.  (Note that we always take $\eta\gg\eta_R$; that is, we assume the GWs are observed significantly after the MD epoch has ended.)  This decomposition is primarily useful for analytically performing the oscillation average in eq.~\eqref{eq:spectra}.  We perform the matching at the time $\eta_G$ defined as
\begin{equation}
\label{eq:eta_G_def}
    k^2=10^{-12}\frac{a''(\eta_G)}{a(\eta_G)}  ,
\end{equation}
which is motivated by the observation from eq.~\eqref{eq:greens} that the RD limit is valid when $k^2\gg a''(\eta)/a(\eta)$. Thus, when $\Bar{\eta}>\eta_G$ we simply use the RD Green's function
\begin{equation}
    G_k^{\rm RD}(\eta,\Bar{\eta})=\frac{1}{k}\sin{(k\eta-k\Bar{\eta})}.
\end{equation}

Our results below indicate that the abrupt change in the piecewise definitions of the Green's function produces unphysical oscillations in the signal, and we note that this matching necessarily introduces similar abrupt changes.  We have defined $\eta_G$ in eq.~\eqref{eq:eta_G_def} so that the resulting oscillations induced in the spectra are negligible.  

For each value of $k$ we sample and interpolate $G$ over $\Bar{\eta}$ densely enough so that the integral in eq.~\eqref{eq:I} can be performed accurately.  Since the Green's function depends only on $k$, and not $u$ or $v$, we can reuse the calculation for each point sampled in the integrals of eq.~\eqref{eq:spectra}.

The predominant computational complexity in calculating the induced GWs comes from evolving the gravitational potential.  This involves solving a coupled differential equation for the gravitational potential, energy density perturbations $\delta_{m/r}$, and velocity divergences $\theta_{m/r}$ of matter and radiation respectively.  To allow for as direct a comparison as possible with references~\cite{Inomata:gradual,Inomata:sudden} we choose to evolve the perturbations in the Newtonian gauge.  This causes complications for a time dependent decay rate, as the presence of $\Phi$ in \eqref{eq:metric} causes time to not be synchronised with the matter.  Reference~\cite{Inomata:pbh} deals with this by solving for the perturbations in the synchronous gauge, with coordinates comoving with the matter, before transforming to the Newtonian gauge to calculate the induced gravitational wave spectrum.  We instead transform the synchronous gauge equations in~\cite{Poulin:2016nat} to the Newtonian gauge while including the time dependence of the decay rate.  Neglecting anisotropic stress of radiation, the perturbations for a given wavenumber $k$ evolve as
\begin{align}
    \label{eq:deltam}
    \delta_m'&=-\theta_m+3\Phi'-a\Gamma(\eta)\Phi-a\Gamma'(\eta)\frac{\theta_m}{k^2} \\
    \theta_m'&=-\mathcal{H}\theta_m+k^2\Phi \\
    \delta_r'&=-\frac{4}{3}(\theta_r-3\Phi')+a\Gamma(\eta)\frac{\rho_m}{\rho_r}(\delta_m-\delta_r+\Phi)+a\Gamma'(\eta)\frac{\rho_m}{\rho_r}\frac{\theta_m}{k^2}\\
    \label{eq:thetar}
    \theta_r'&=\frac{k^2}{4}\delta_r+k^2\Phi-a\Gamma(\eta)\frac{3\rho_m}{4\rho_r}\left(\frac{4}{3}\theta_r-\theta_m\right)\\
    \label{eq:phiprime}
    \Phi'&=-\frac{k^2\Phi+3\mathcal{H}^2\Phi+\frac{3}{2}\mathcal{H}^2\left( \displaystyle{\frac{\rho_m\delta_m+\rho_r\delta_r}{\rho_{\rm tot}}} \right)}{3\mathcal{H}}
\end{align}
where the additional $\Gamma'$ terms arise in the transition to Newtonian gauge, due to the non-trivial time-dependence of $\Gamma$.  These terms were not included in \cite{Inomata:sudden}, although as noted above a time-dependent decay rate is necessary to have a transition with $\Gamma \gg \mathcal{H}$.  The derivation of these equations may be found in appendix~\ref{app:perturbations}.  We evolve these equations forward in time from the initial conditions
\begin{equation}
    \delta_{m,0}=-2\Phi_0,\;\;\delta_{r,0}=\frac{4}{3}\delta_{m,0},\;\;\theta_{m,0}=\theta_{r,0}=\frac{k^2\eta}{3}\Phi_0.
\end{equation}
By taking $\Phi_0=1$ we effectively solve for the transfer function.  

We compare solving this system of equations numerically to the approximations used in the literature.  First, in the gradual case~\cite{Inomata:gradual}, the oscillatory period was neglected as the gravitational potential decayed significantly.  In place of a numerical solution, the following fit, which is $k$ independent, was used
    \begin{equation}
        \Phi_{\rm fit}=\begin{cases}
        \exp\left(-\tfrac{2}{3}\left(\tfrac{\eta}{\eta_R}\right)^3\right) & \eta<\eta_R \\
        \exp\left(-2\left(\left(\tfrac{\eta}{\eta_R}\right)^2-\tfrac{\eta}{\eta_R}+\tfrac{1}{3}\right)\right) & \eta\geq\eta_R.
    \end{cases}
    \end{equation}
Because we include the oscillating tail, our analysis is not $k$ independent and must be solved for each wave number, which increases the computational time.  This is especially pronounced because if the $k$ dependence is neglected, then $f$ and $I$ become independent of $u$ and $v$.  Consequently, $I$ factors out of the integral in eq.~\eqref{eq:spectra}.

For the sudden case~\cite{Inomata:sudden}, conversely, the gravitational potential experiences no decay, instantaneously switching from a constant solution during the MD era to an oscillating solution with large amplitude at $\eta_R$.  In this case the spectrum is almost entirely sourced by the RD component and we will see below that an analytic solution exists both for $\Phi$ and $I$ during the RD era.  This means that their analysis completely avoids the expensive step of evolving the perturbations and the remaining integral for the power spectrum eq.~\eqref{eq:spectra} depends only on analytic functions of $u$ and $v$.

\begin{figure}[tbp]
    \centering
    \includegraphics[width=\figurewidth]{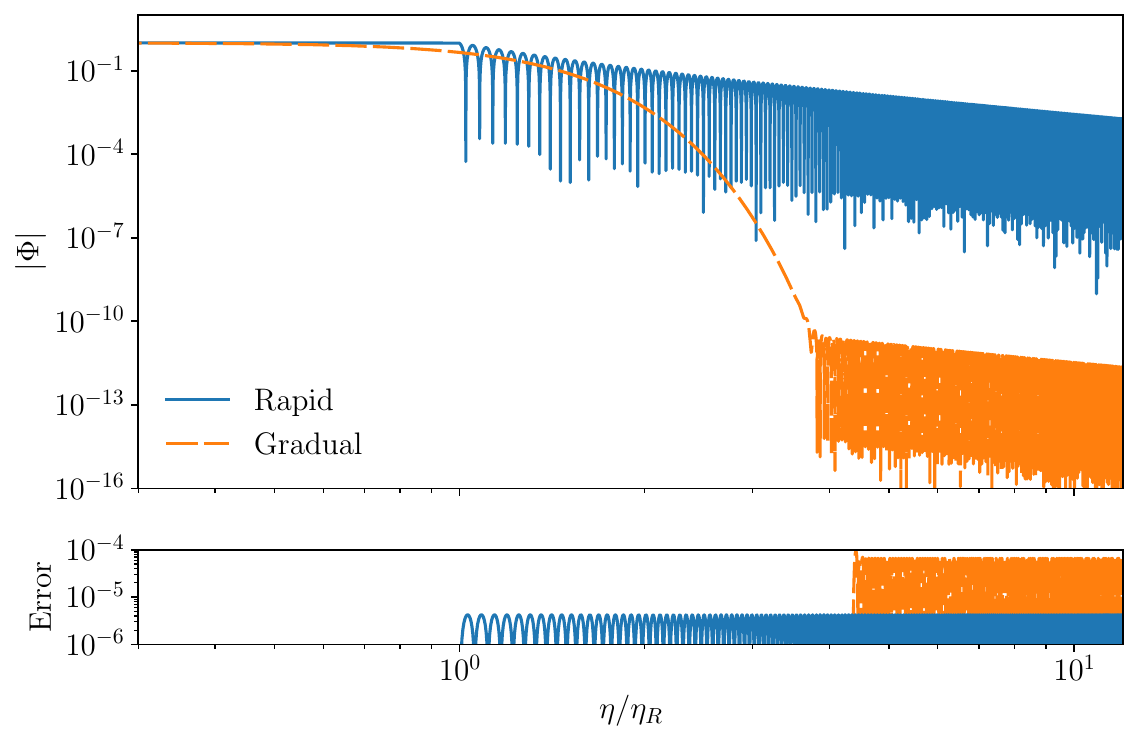}
    \caption{Time dependence of the gravitational potential $\Phi$ for wavenumber $k=100/\eta_R$.  The evolution for a rapid transition with $\beta=3000/\eta_R$ in blue is contrasted with that of a gradual transition with $\beta=0/\eta_R$ in dashed orange.  The error is that of the analytic approximation during the RD era defined as $|\Phi-\Phi_A|/\Phi_{\rm amp}$ with $\Phi_{\rm amp}$ as defined in eq.~\eqref{eq:amplitude}.  We note that $\eta_{\rm osc}$ differs in the rapid and graduate cases, explaining why the error appears at different times.}
    \label{fig:phi}
\end{figure}

Our approach, as we describe below, is to numerically solve for the gravitational potential $\Phi$ before matching onto a late time analytic solution.  In figure~\ref{fig:phi} we show the time dependence of the gravitational potential, calculated numerically, for wavenumber $k=100/\eta_R$ in the case of a rapid transition in blue and a gradual transition in dashed orange.  As expected, during the MD era $\Phi$ remains constant before experiencing some period of decay and then beginning to oscillate.  The amount of decay depends on how rapid the transition is, as we see that in the rapid limit the potential experiences no decay, whereas in the gradual limit it decays by roughly ten orders of magnitude.  In both cases, the potential continues to oscillate and decays at a slower rate during the RD era.  

The oscillations during the RD era complicate the numerical integration over $\Bar{\eta}$ in eq.~\eqref{eq:I}.  These oscillations are crucial in the rapid scenario, as they give rise to the resonant poltergeist peak~\cite{Inomata:sudden}.  To capture this behaviour we numerically evolve the perturbations during the transition, but once $\delta_r$ comes to dominate the last term in eq.~\eqref{eq:phiprime} we match onto an analytic solution.  We define the time $\eta_{\rm osc}$ at which we match onto the analytic solution to be when
\begin{equation}
    \frac{\rho_m\delta_m}{\rho_r\delta_r}=10^{-4}
\end{equation}
and as such it will depend both on $k$ and the suddenness of the transition $\beta$.  

We now describe the analytic solution that we match onto.  It makes use of the fact that, away from the transition, the evolution of $\Phi$ in a single component fluid is described by~\cite{GorbunovRubakov}
\begin{equation}
    \Phi''+3(1+w)\mathcal{H}\Phi'+wk^2\Phi=0,
\end{equation}
where, in this case, $\mathcal{H}$ is evaluated from the analytic approximation in eq.~\eqref{eq:scalefactor}, as we are away from the transition.  During the RD era this has the solution
\begin{equation}
    \Phi_A(k,\eta)=A(k,\eta_{\rm osc})\mathcal{J}(k\eta)+B(k,\eta_{\rm osc})\mathcal{Y}(k\eta)
\end{equation}
where $\mathcal{J}$ and $\mathcal{Y}$ are defined in terms of the spherical Bessel functions, $j_1(x)$ and $y_1(x)$,
\begin{align}
    \mathcal{J}(k\eta)=\frac{3\sqrt{3}\;j_1\left(\frac{k\eta-k\eta_{\rm RD}/2}{\sqrt{3}}\right)}{k\eta-k\eta_{\rm RD}/2}\\
    \mathcal{Y}(k\eta)=\frac{3\sqrt{3}\;y_1\left(\frac{k\eta-k\eta_{\rm RD}/2}{\sqrt{3}}\right)}{k\eta-k\eta_{\rm RD}/2},
\end{align}
where we have defined $\eta_{RD}$ above.  The coefficients $A(k,\eta_{\rm osc})$ and $B(k,\eta_{\rm osc})$ are given by imposing continuity of the solution and its derivative at $\eta_{\rm osc}$.

We see that during the RD era the amplitude of the potential decays as
\begin{equation}
\label{eq:amplitude}
    \Phi_{\rm amp}=\frac{9\sqrt{A(k,\eta_{\rm osc})^2+B(k,\eta_{\rm osc})^2}}{(k(\eta-\eta_{\rm RD}/2))^2}
\end{equation}
which we use to normalise the error in the analytic solution as $|\Phi-\Phi_A|/\Phi_{\rm amp}$, which is shown in the lower panel of figure~\ref{fig:phi}.  We only show the error for $\eta>\eta_{\rm osc}$ when we actually use the analytic solution, where $\eta_{\rm osc}\approx 4.4\eta_R$ in the gradual case and $\eta_{\rm osc}\approx \eta_R$ in the rapid. While we enforce a relative error of $10^{-11}$ in the ODE solver, we find that the error in the approximation is significantly larger than this, for the gradual case being of order $10^{-4}$.  This is because we neglect the term $\rho_m\delta_m$ in equation~\eqref{eq:phiprime} which is of order $10^{-4}$ times the dominant term $\rho_r\delta_r$ at $\eta_{\rm osc}$ by definition.  The error is much better in the rapid case as the $\rho_m\delta_m$ term simply decays away faster due to the sharp turn on of the decay rate.

There is a balance between setting $\eta_{\rm osc}$ late enough that the $\rho_m\delta_m$ term is truly negligible but also early enough to avoid spending too much time accurately integrating the ODE for an oscillating solution.  Since the main purpose for transitioning to an analytic expression is evaluating the integral for $I$ in eq.~\eqref{eq:I}, the error in the amplitude translates directly to the error in $I$.  The reduced accuracy in the gradual case is offset by the more than ten orders of magnitude suppression by the time oscillations begin, making their contribution to the integral negligible.

The reasoning for using the analytic expression at late times is two-fold.  Firstly, as discussed above, finding an accurate numerical solution for $\Phi$ during the oscillations is computationally expensive.  Utilising the analytic expression takes roughly a quarter of the time it would take to numerically integrate the ODE over the time ranges shown in figure~\ref{fig:phi}.  Secondly, due to the oscillations, it would be difficult to perform the integral for $I$, in eq.~\eqref{eq:I}, numerically.  The analytic expression $\Phi_A$ admits an analytic integral for $I$.  We evaluated this integral (using {\tt Mathematica}) and, after transforming the integration variable to $\bar{x}=k(\bar{\eta}-\eta_{\rm RD}/2)$ we found our result is consistent with that of appendix~A of ref.~\cite{Kohri:2018awv} making the substitutions
\begin{align}
    A(u)&=\frac{9}{10}A(uk,\eta_{\rm osc})\\
    B(u)&=\frac{9}{10}B(uk,\eta_{\rm osc})\\
    C&=-\cos\left(k(\bar{\eta}-\eta_{\rm RD}/2)\right)\\
    D&=\sin\left(k(\bar{\eta}-\eta_{\rm RD}/2)\right)\\
    x_1&=k(\eta_{\rm osc}-\eta_{\rm RD}/2)
\end{align}
and taking the limit $x_2\rightarrow\infty$.  The factors of $9/10$ in $A$ and $B$ are due to ref.~\cite{Kohri:2018awv} having normalised $\Phi$ to unity during the RD era.

Finally, once we have solved for $I$, we find the GW signal by performing the integrals in eq.~\eqref{eq:spectra}.  The details of this integration can be found in appendix~\ref{app:code}, along with other details on our numerical techniques.  We note that in this final integral we allow up to 5 percent error, as this is sufficient precision to compare with the gradual and rapid limits from previous work and make qualitative statements about the spectra and potential detectability.

It is useful for comparing back to phenomenological scenarios to derive a relationship between conformal time $\eta$ and temperature $T$ valid during the RD era~\cite{Inomata:2018epa}.  From the Friedmann equations we have
\begin{equation}
    \frac{aH}{a_{\rm eq}H_{\rm eq}}=\frac{a}{a_{\rm eq}}\sqrt{\frac{\rho}{2\rho_{\gamma,{\rm eq}}}}
\end{equation}
where the subscript eq denotes evaluation at the late matter radiation equality time and $\rho_{\gamma}$ is the photon energy density.  Near the reheating transition it is adequate to use eq.~\eqref{eq:scalefactor} which, during the RD era, gives $aH = (\eta-\eta_R/2)^{-1}$.  Assuming entropy conservation $g_{s,\rm eq}a_{\rm eq}^3 T_{\rm eq}^3=g_s a^3 T^3$ where $g_s$ is the number of relativistic degrees of freedom of entropy.  We reduce the ratio of energy densities by taking $\rho\propto gT^4$ where $g$ is the effective number of relativistic degrees of freedom of radiation.  Taking $a_{\rm eq}H_{\rm eq}=k_{\rm eq}=0.01041\; {\rm Mpc^{-1}}$ \cite{Planck:2018vyg,Planck:2015fie} gives
\begin{equation}
    \frac{(\eta-\eta_R/2)^{-1}}{0.01041\; {\rm Mpc^{-1}}}=\frac{1}{\sqrt{2}}\left(\frac{g_{s,\rm eq}}{g_s}\right)^{1/3} \left(\frac{g}{g_{\rm eq}}\right)^{1/2}\frac{T}{T_{\rm eq}}.
\end{equation}
We calculate $T_{\rm eq}=T_0(1+z_{\rm eq})=0.801\;{\rm eV}$ from the current CMB temperature $T_0=2.7255\; {\rm K}$  \cite{2009ApJ...707..916F} and the redshift $z_{\rm eq}=3411$ \cite{Planck:2018vyg}.  Finally taking $g_{\rm eq}=3.38$ and $g_{s,{\rm eq}}=3.94$ \cite{Husdal:2016haj} we obtain
\begin{equation}
    \eta-\frac{\eta_R}{2}=58.2\;{\rm MeVpc} \left(\frac{g_s}{106.75}\right)^{1/3} \left(\frac{g}{106.75}\right)^{-1/2}T^{-1}.
\end{equation}
We use this both to evaluate the reheating temperature, as well as to match our evolution of the Friedmann equations, eqs.~\eqref{eq:fried_m} to \eqref{eq:fried_a}, onto the standard RD era for $\eta\gg\eta_R$.

After the gravitational potential source decays away enough, the GW energy density parameter $\Omega_{\rm GW}$ comes to be constant during the RD era.  For reference, we define the time at which $\Omega_{\rm GW}$ becomes constant to be $\eta_c$.  The GWs will evolve further during the late MD era with the present value of the energy density parameter given by~\cite{Inomata:sudden}
\begin{equation}
    \Omega_{\rm GW}(\eta_0,k)=0.39\left(\frac{g(\eta_c)}{106.75}\right)^{-1/3}\Omega_{r,0}\Omega_{\rm GW}(\eta_c,k)
\end{equation}
where $\Omega_{r,0}$ is the present value of the radiation energy density parameter and $\eta_0$ is the current conformal time.  This also takes into account the change in the effective relativistic degrees of freedom $g$ between when the GWs were sourced and now.

\section{Results}
\label{sec:results}

In this section, we will present the results of our improved calculation.

\begin{figure}[tbp]
    \centering
    \includegraphics[width=\figurewidth]{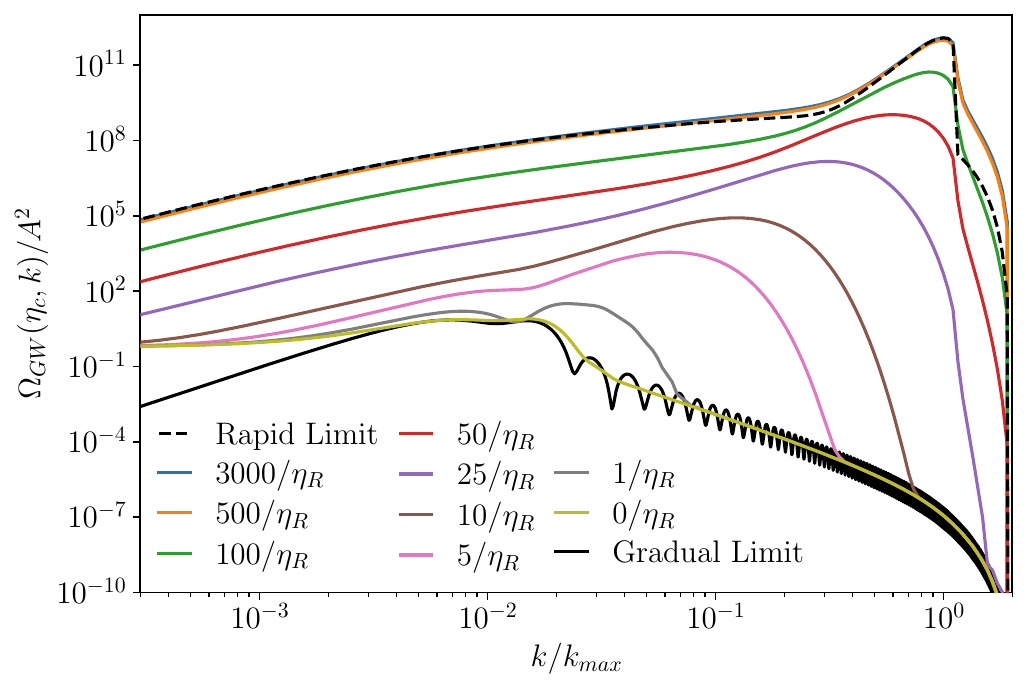}
    \caption{GW power spectrum induced from a scale invariant power spectrum ($n_s=1$).  We fixed $\eta_{\Gamma}$ in eq.~\eqref{eq:decay} to produce a transition occurring at $T_R = 100 \; {\rm GeV}$.  The curves are labelled by the value of $\beta$.  We compare our results to the rapid limit from ref.~\cite{Inomata:sudden} in dashed black and the gradual limit from ref.~\cite{Inomata:gradual} in solid black.}
    \label{fig:spectra}
\end{figure}

\begin{figure}[tbp]
    \centering
    \includegraphics[width=\figurewidth]{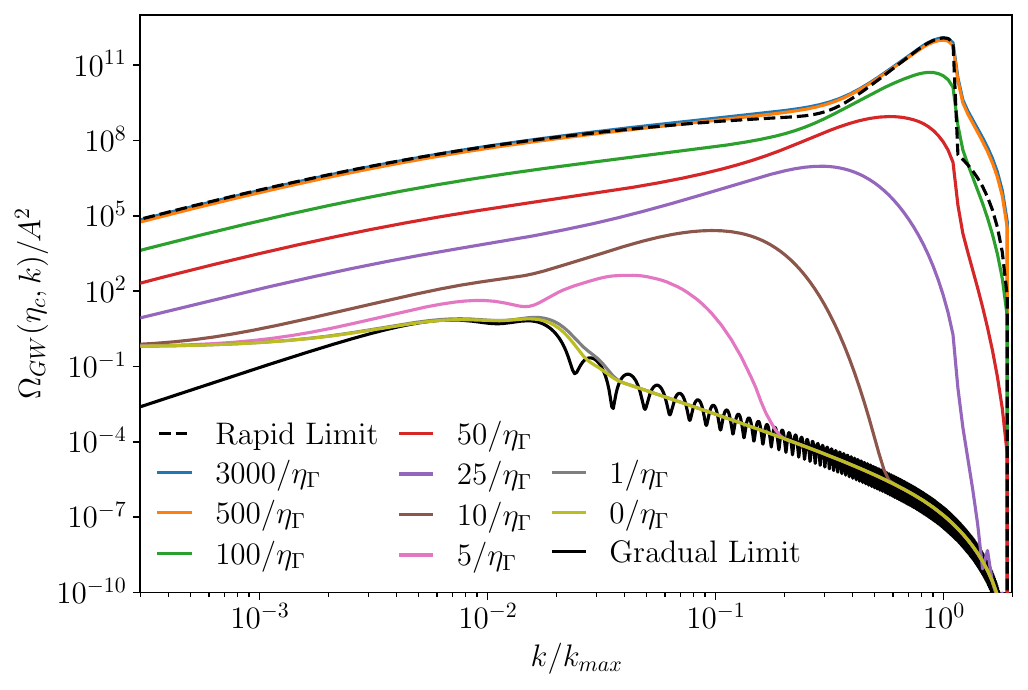}
    \caption{GW power spectrum induced from a scale invariant power spectrum ($n_s=1$).  We fixed $\eta_{\Gamma}=1.164\times10^{-9} \; {\rm Mpc}$ in eq.~\eqref{eq:decay} so that the most rapid transition occurs at $T_R = 100 \; {\rm GeV}$.  The curves are labelled by the value of $\beta$.  We compare our results to the rapid limit from ref.~\cite{Inomata:sudden} in dashed black and the gradual limit from ref.~\cite{Inomata:gradual} in solid black.}
    \label{fig:spectra_decay}
\end{figure}

Our main results are presented in figures~\ref{fig:spectra} and \ref{fig:spectra_decay}, which show our numerical calculations of the GW spectra induced from a scale invariant curvature spectrum for a variety of matter-radiation transition speeds.  In figure~\ref{fig:spectra}, for each value of $\beta$ we we set $\eta_{\Gamma}$ such that the transition occurs at $\eta_R=1.164\times10^{-9} \; {\rm Mpc}$, corresponding to a reheating temperature $T_R = 100 \; {\rm GeV}$.  We also set $\Gamma_{\rm max}=500 \Gamma_R$, where $\Gamma_R = 1.590\times10^{-14} \; {\rm GeV}$ is the constant decay rate that produces a transition at $\eta_R$, to ensure that for large $\beta$ the matter evaporates rapidly when the decay turns on.  In contrast figure~\ref{fig:spectra_decay} demonstrates the same GW spectra for a different choice of the decay rate parameters.  In this case we set  $\Gamma_{\rm max}= 7.948\times10^{-12} \; {\rm GeV}$, the same as before, and fix $\eta_{\Gamma}=1.164\times10^{-9} \; {\rm Mpc}$ so that the rapid limit is the same between the two plots.  Now $\eta_R$ is no longer fixed and as $\beta$ decreases, the transition shifts to earlier times.  The parameters of figures~\ref{fig:spectra} and \ref{fig:spectra_decay} are summarised in tables~\ref{tab:inputs1} and \ref{tab:inputs2} respectively.
\begin{table}[tbp]
    \centering
    \begin{tabular}{cccc}
    \hline
         $\beta$ & $\eta_{\Gamma}$ (Mpc) & $\Gamma_{\rm max}$ (GeV) & $\tau H$ \\
         \hline
         $3000/\eta_R$ & $1.157\times10^{-9}$ & $7.948\times10^{-12}$ & $2.685\times10^{-3}$ \\
         $500/\eta_R$ & $1.158\times10^{-9}$ & $7.948\times10^{-12}$ & $9.926\times10^{-3}$ \\
         $100/\eta_R$ & $1.171\times10^{-9}$ & $7.948\times10^{-12}$ & $4.490\times10^{-2}$ \\
         $50/\eta_R$ & $1.192\times10^{-9}$ & $7.948\times10^{-12}$ & $8.641\times10^{-2}$ \\
         $25/\eta_R$ & $1.242\times10^{-9}$ & $7.948\times10^{-12}$ & $1.623\times10^{-1}$ \\
         $10/\eta_R$ & $1.420\times10^{-9}$ & $7.948\times10^{-12}$ & $3.423\times10^{-1}$ \\
         $5/\eta_R$ & $1.750\times10^{-9}$ & $7.948\times10^{-12}$ & $5.291\times10^{-1}$ \\
         $1/\eta_R$ & $4.612\times10^{-9}$ & $7.948\times10^{-12}$ & $8.062\times10^{-1}$ \\
         $0$ & - & $3.179\times10^{-14}$ & $8.373\times10^{-1}$ \\
         \hline
    \end{tabular}
    \caption{Parameters associated with figure~\ref{fig:spectra}, where we fixed $\eta_R=1.164\times10^{-9} \; {\rm Mpc}$, which corresponds to $T_R=100\;{\rm GeV}$ and $k_{\rm max}=3.867\times 10^{11}\;{\rm Mpc^{-1}}$.  When $\beta=0/\eta_R$ this uniquely determines $\Gamma_{\rm max}$.  In the opposite limit, to have a rapid transition, it is necessary to have $\Gamma_{\rm max} \gg H$ and thus we use larger values for nonzero $\beta$.}
    \label{tab:inputs1}
\end{table}

\begin{table}[tbp]
    \centering
    \begin{tabular}{ccccc}
    \hline
         $\beta$ & $\eta_R$ (Mpc) & $T_R$ (GeV)& $k_{\rm max}\;({\rm Mpc^{-1}})$ & $\tau H$ \\
         \hline
         $3000/\eta_{\Gamma}$ & $1.170\times10^{-9}$ & $99.42$ & $3.844\times10^{11}$ & $2.663\times10^{-3}$ \\
         $500/\eta_{\Gamma}$ & $1.169\times10^{-9}$ & $99.52$ & $3.848\times10^{11}$ & $9.877\times10^{-3}$ \\
         $100/\eta_{\Gamma}$ & $1.157\times10^{-9}$ & $100.6$ & $3.890\times10^{11}$ & $4.516\times10^{-2}$ \\
         $50/\eta_{\Gamma}$ & $1.136\times10^{-9}$ & $102.5$ & $3.962\times10^{11}$ & $8.837\times10^{-2}$ \\
         $25/\eta_{\Gamma}$ & $1.087\times10^{-9}$ & $107.1$ & $4.141\times10^{11}$ & $1.722\times10^{-1}$ \\
         $10/\eta_{\Gamma}$ & $9.237\times10^{-10}$ & $126.0$ & $4.873\times10^{11}$ & $4.003\times10^{-1}$ \\
         $5/\eta_{\Gamma}$ & $6.637\times10^{-10}$ & $175.3$ & $6.793\times10^{11}$ & $6.617\times10^{-1}$ \\
         $1/\eta_{\Gamma}$ & $1.387\times10^{-10}$ & $838.9$ & $3.215\times10^{12}$ & $8.300\times10^{-1}$ \\
         $0$ & $7.400\times10^{-11}$ & $1572$ & $6.048\times10^{12}$ & $8.259\times10^{-1}$ \\
         \hline
    \end{tabular}
    \caption{Parameters associated with figure~\ref{fig:spectra_decay}, where we fixed $\eta_{\Gamma}=1.164\times10^{-9} \; {\rm Mpc}$ and $\Gamma_{\rm max}=7.948\times10^{-12}\;{\rm GeV}$.}
    \label{tab:inputs2}
\end{table}

In addition to characterising the transition in terms of $\beta$ we can also compare the duration in regular time, $\tau$, directly to the Hubble rate at matter radiation equality, $H(\eta_{\rm eq})$.  This gives us a more intuitive quantity to compare real models to, for example, in a gradual transition $\tau H\sim 1$ whereas a phase transition can complete with $\tau H \in (10^{-4},10^0)$.  We can compute $\tau$ from the conformal time
\begin{equation}
    \tau = \int_{\eta_i}^{\eta_f}d\overline{\eta} \; a(\overline{\eta}).
\end{equation}
We define the initial and final times $\eta_i$ and $\eta_f$ from the comoving matter energy density $\rho_m(\eta)(a(\eta)/a(\eta_{\rm MD}))^3$, where $\eta_{\rm MD}$ is the start of the MD era.  The start of the transition $\eta_i$ is defined as when one percent of the matter has decayed and the end $\eta_f$ is when only a factor of $1/e$ remains.

Figures~\ref{fig:spectra} and \ref{fig:spectra_decay} are motivated by different cosmological model-building motivations: if a model has heavy matter whose decay is uncertain, then figure~\ref{fig:spectra_decay} may be more useful.  On the other hand, if one is building a model by imposing a transition at a specific time, adjusting the type or amount of heavy matter, then figure~\ref{fig:spectra} may be most useful.  Both figures share the same qualitative features, however, demonstrating how the signal evolves between the slow and fast limits.

For comparison, the dashed, black curves in both figures show the rapid limit, constructed from the analytic approximation to the large scale (small $k\slash k_{\rm max}$) and resonant peak contributions to the spectrum, derived in the appendix of ref.~\cite{Inomata:sudden}.  Similarly, the solid, black curve denotes the gradual limit reconstructed from ref.~\cite{Inomata:gradual} using their approximate fits to the scale factor, equation of state and gravitational potential.  We see excellent agreement with our numerical code in the rapid limit with the nearly coincident blue and orange curves ($k=3000/\eta_R$ and $k=500/\eta_R$ respectively) perfectly matching the analytic approximation in the regions for which it is valid, both in the broad low $k \slash k_{\rm max}$ regime of the spectrum, as well as at the resonant peak.  For the short scale part of the spectrum (large $k\slash k_{\rm max}$) away from the peak, they diverge from the approximate rapid limit because our result is a complete computation of the spectrum.  This is consistent with the numerical results from ref.~\cite{Inomata:sudden} which similarly diverge from the approximation in this region.  We include both the orange $\beta=500/\eta_R$ curve and the blue $\beta=3000/\eta_R$ curve to demonstrate that by $\beta=500/\eta_R$ we have effectively converged to the rapid limit.  The agreement is less precise in the gradual case, with the olive curve ($k=0$) matching the overall scaling of the dashed curve for large $k$, but differing at small $k$.  We also see that our analysis does not have the oscillations at large $k$ values.  Both of these will be discussed below.

A noteable difference between the curves plotted in figure~\ref{fig:spectra} and figure~\ref{fig:spectra_decay} is that when $\eta_{\Gamma}$ is fixed, $\eta_R$ varies, and thus the time scale at which the transition occurs changes.  Similarly, $k_{\rm max}$ is different for each curve and the spectrum is shifted to higher frequencies for earlier $\eta_R$.  This dependence of $\eta_R$ and $k_{\rm max}$ on $\beta$ is summarised in table~\ref{tab:inputs2}.  However, since the spectrum is a function of $k \slash k_{\rm max}$ this is not immediately evident from figure~\ref{fig:spectra_decay}.

The most important feature of these results is demonstrating the evolution of the resonant-like peak (the so-called poltergeist effect) as the matter-radiation transition time increases.  First, we see that the peak does exist for finite $\beta$, and therefore it is not a numerical artefact of any of the step functions used in the analyses of ref.~\cite{Inomata:sudden}.  We see that as the duration of the transition increases, the entire spectrum becomes suppressed as would be expected.  In fact we see that the red curve (corresponding to $\beta=50/\eta_R$) is already suppressed by roughly three orders of magnitude compared to the rapid limit.  This corresponds to a transition duration of roughly $\tau H=8.64\times10^{-2}$, demonstrating that to achieve the theoretical maximum strength signal, the transition needs to be faster than $\mathcal{O}(0.01)$ of the Hubble time.

In addition to this, the resonant peak broadens, which can be seen particularly in the $\beta = 3000 \slash \eta_R$ and $\beta = 500 \slash \eta_R$ lines, which are nearly coincident to the right of the resonant peak.  The resonance can be understood as the amplification of GWs that are comoving with the sound waves produced in the plasma after the sudden transformation of matter into radiation.  When the transition is spread out over a small time, the sound waves exist in a frequency band, amplifying a range of comoving GWs and broadening the peak.  However, if the transition time is sufficiently long, then the overdensities dissipate instead of sourcing sound waves.

\begin{figure}[tbp]
    \centering
    \includegraphics[width=\figurewidth]{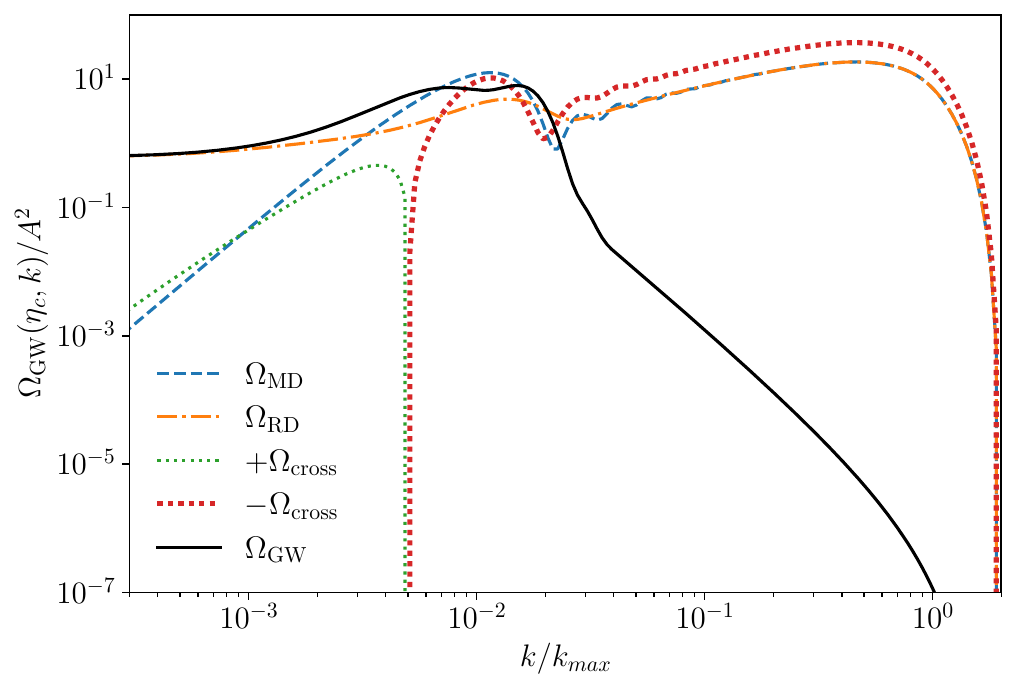}
    \caption{The GW signal $\Omega_{\rm GW}$ for a gradual transition with $\beta=0/\eta_R$ decomposed into contributions from the MD and RD eras and the cross term.}
    \label{fig:gradualcomps}
\end{figure}

Next we discuss the gradual limit, and in particular, the differences between our analysis and ref.~\cite{Inomata:gradual} that we mentioned above.  For $\beta\lesssim 10/\eta_R$ (the brown curve and below) we see that for large enough $k \slash k_{\rm max}$ the spectra converges to the gradual limit.  The significant suppression at large $k$ was previously noted in the literature~\cite{Inomata:gradual}, which has been understood as arising from the cross term in $I^2$ when $I$ is broken into MD and RD contributions,
\begin{align}
     I(u,v,k,\eta)\nonumber &= k\int _0 ^{\eta_R} d \bar{\eta}\; \frac{a(\Bar{\eta})}{a(\eta)} k G_k ^{ {\rm MD} } (\eta , \bar{\eta})f(u,v,k,\Bar{\eta}) \nonumber + k\int _{\eta_R} ^{\eta} d\bar{\eta}\; \frac{a(\Bar{\eta})}{a(\eta)} k G^{\rm RD} _k (\eta , \bar{\eta}) f(u,v,k,\Bar{\eta}) \nonumber \\
    &= I_{{\rm MD}} (u,v,k,\eta) + I_{\rm RD}(u,v,k,\eta)  .
\end{align}
The GW spectrum itself can then be decomposed into a term from the RD era (sourced by $\overline{I_{\rm RD}^2}$), a term from the MD era (sourced by $\overline{I_{\rm MD}^2}$) and a cross term (sourced by $2\overline{I_{\rm MD}I_{\rm RD}}$), that is
\begin{equation}
\label{eq:split}
    \Omega _{\rm GW} = \Omega _{\rm RD} + \Omega _{\rm MD} + \Omega _{\rm cross} \ .
\end{equation}
Since we numerically evolve all relevant quantities, and therefore don't use piecewise expressions for the scale factor, Green's function and gravitational potential, we do not need to decompose the signal into these components.  However, to facilitate comparison to ref.~\cite{Inomata:gradual}, we have identified the corresponding contributions in our analysis (using $\beta=0/\eta_R$), and the results are shown in figure~\ref{fig:gradualcomps}.

We see that for $k\gtrsim0.03 k_{\rm max}$ each contribution is comparable in magnitude with $\Omega_{\rm MD}\simeq\Omega_{\rm RD}\simeq-\Omega_{\rm cross}/2$ such that the resulting signal is many orders of magnitude smaller than any individual contribution.  To understand why the cross term cancels so precisely, we note that in the large $k \slash k_{\rm max} $ regime, the gravitational potential undergoes significant decay before it oscillates with suppressed amplitude.  Therefore, the contribution from the RD era $\Omega_{\rm RD}$ is nearly entirely sourced during the regime in which $\Phi$ is rapidly decreasing, before it even begins to oscillate.  This is also the regime with the most significant contribution to the mixed term, as the matter density has not entirely disappeared yet.

This cancellation particularly demonstrates the importance of numerical precision in these calculations, and as can be seen in figures \ref{fig:spectra}  and \ref{fig:spectra_decay} our results in the large $k \slash k_{\rm max}$ regime are in agreement with version 3 of ref.~\cite{Inomata:gradual}\footnote{We thank the authors of ref.~\cite{Inomata:gradual} for updating their analysis after discussion.}.

\begin{figure}[tbp]
    \centering
    \includegraphics[width=\figurewidth]{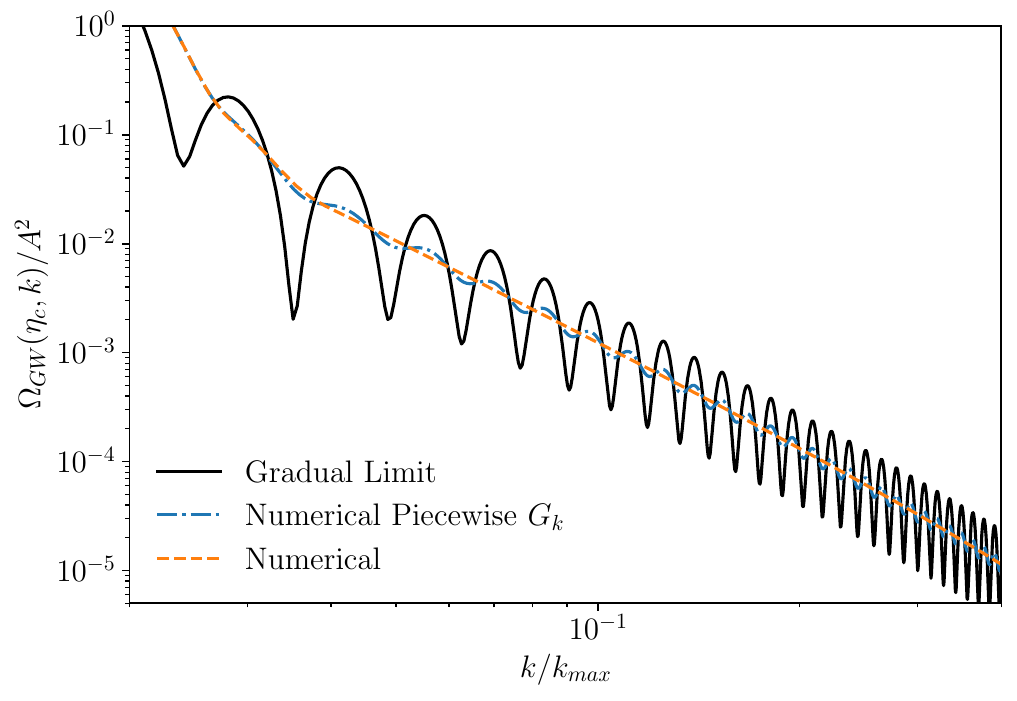}
    \caption{GW power spectrum comparing the gradual limit for different levels of approximation.  The solid black curve shows the limit reproduced from ref.~\cite{Inomata:gradual} with an instantaneous approximation for $a$ and the Green's function.  The dash-dotted blue curve show our numerical result if we were to use the same piecewise approximation for the Green's function as in ref.~\cite{Inomata:gradual}.  Lastly the dashed orange curve shows our fully numerical results.}
    \label{fig:GradualGreensCompare}
\end{figure}

Quite crucially though, once the power spectrum has converged to the gradual limit it does not oscillate with $k$ as it does in the results from ref.~\cite{Inomata:gradual}.  We observe that the oscillations in $k$ have a period of $2\pi/\eta_R$, indicating their origin is from the matching between the MD and RD eras.  These oscillations arise both from discontinuities in the second derivatives of $a$ and $f$ as well as from the piecewise treatment of the Green's function.  This can be seen in figure~\ref{fig:GradualGreensCompare}, which shows that if we use our numerical solution for the scale factor in place of eq.~\eqref{eq:scalefactor} (but retain the piecewise approximation for the Green's functions) the oscillations are suppressed, and then if we further replace the approximate Green's function with our numerical solution they are absent.  As such, our signals are free of the oscillatory artefact and highlight the true characteristic of the signal from a gradual transition.

As we mentioned, we also find that our numerical results differ considerably from the gradual limit for small $k \slash k_{\rm max}$, converging to the olive $\beta=0/\eta_R$ curve.  This is due to the gradual limit employing a fit to $\Phi$ valid for $k\gtrsim 30/\eta_{\rm eq}$, where $\eta_{\rm eq}$ is the matter-radiation equality time of the transition in question (see eq.~(3.10) and following discussion in ref.~\cite{Inomata:gradual}).  By numerically evolving $\Phi$ we do not face the same restriction and are able to account for oscillations in $\Phi$ during the MD era that become large for small $k$.  This is evident in figure~\ref{fig:gradualcomps} where we see that $\Omega_{\rm RD}$ comes to dominate the spectrum for small $k$.

We also observe that for finite-duration transitions, the GW signal at large $k \slash k_{\rm max}$ converges to the slow transition regime faster than the GW signal at lower $k \slash k_{\rm max}$.  This is also a consequence of the gravitational potential undergoing significant decay before it oscillates with suppressed amplitude at large $k \slash k_{\rm max} $, which as explained above leads to the precise cancellation between components.  Physically, at these scales the matter overdensities have time to dissipate during the extended matter-radiation transition period and do not source the sound waves that resonantly enhance the GWs.

\begin{figure}[tbp]
    \centering
    \includegraphics[width=\figurewidth]{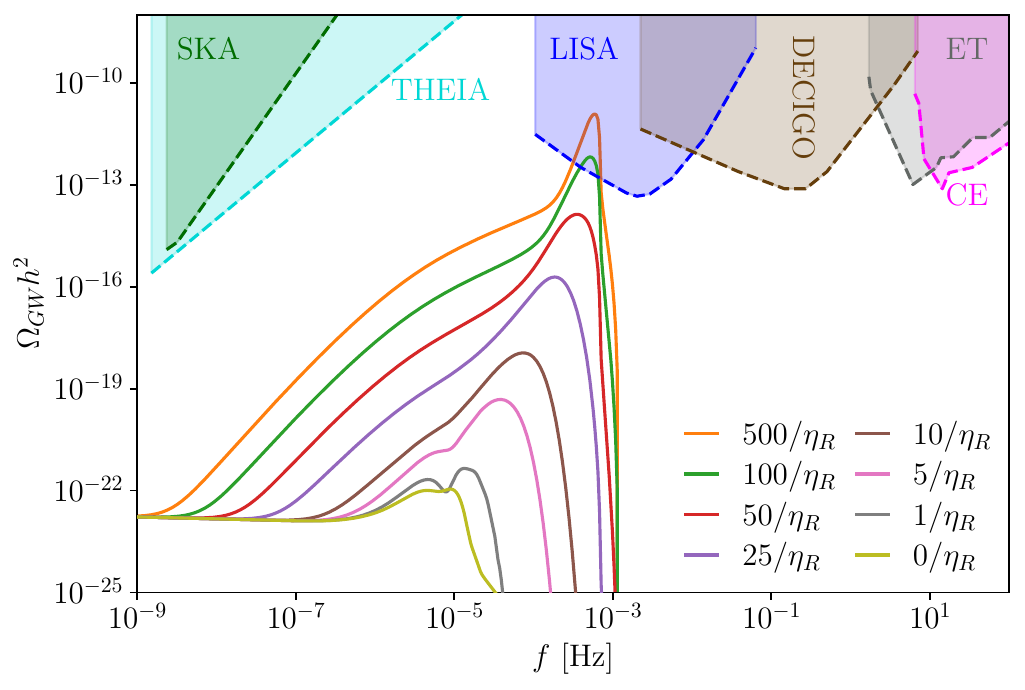}
    \caption{The GW signal induced from a transition at $T_R = 100 \; {\rm GeV}$ with a realistic curvature power spectrum as in eq.~\eqref{eq:curvature}.  The signals are labelled by the value of $\beta$.  We see only the $\beta=500/\eta_R$ and $\beta=100/\eta_R$ spectra would be detectable at LISA.  We compare to sensitivity curves of future GW experiments for LISA observing for four years~\cite{Caprini:2019egz}, the Einstein Telescope observing for one year~\cite{Moore:2014lga,Maggiore:2019uih,Inomata:2018epa}, the Cosmic Explorer~\cite{Reitze:2019iox}, DECIGO with three units observing for one year~\cite{Moore:2014lga,Kawamura:2020pcg}, THEIA observing for twenty years~\cite{thetheiacollaboration2017theia,Garcia-Bellido:2021zgu} and SKA~\cite{Janssen:2014dka}.}
    \label{fig:spectra_sensitivities}
\end{figure}

In figure~\ref{fig:spectra_sensitivities} we show the GWs signals induced during transitions at $T_R=100\;{\rm GeV}$ of varying speeds by a realistic curvature spectrum as in eq.~\eqref{eq:curvature} with $A_s=2.1\time10^{-9}$ and $n_s=0.97$ \cite{Planck:2018vyg}.  We contrast these with the sensitivities of upcoming GW experiments.  We see that for a transition at $100\;{\rm GeV}$, LISA is the applicable detector and would only be sensitive to $\beta\gtrsim 100/\eta_R$ which correspond to transitions with duration $\tau H\gtrsim4.490\times 10^{-2}$.  Since the tilt of the curvature power spectrum is very close to one, as we move to different reheating temperatures, the peak frequency will change but the amplitude of the spectrum will remain fairly constant.  Additionally changes in the effective degrees of freedom $g_c$ will also influence the amplitude of the spectrum, but this is only a small effect, scaling by a factor of $\mathcal{O}(1)$ at most.  As such, only SKA and THEIA would be able to probe more gradual transitions with $\beta \gtrsim50/\eta_R$ and $\tau H\gtrsim 8.641\times 10^{-2}$ occurring with reheating temperatures around $T_R\sim1\;{\rm MeV}$.  Other interferometers will perform similarly to LISA, being sensitive to transitions with duration $\tau H\gtrsim4.490\times 10^{-2}$, with DECIGO probing reheating temperatures around $T_R\sim 30\;{\rm TeV}$ and the Einstein Telescope and Cosmic Explorer probing $T_R\sim 10^6\;{\rm GeV}$.

Finally, we contrast our approach with those of references~\cite{Inomata:pbh,Kasuya:2022cko}.  In these works, the finite width of the transition is included only via a normalisation factor $S$ in the gravitational potential $\Phi$, which describes the suppression of $\Phi$ before its oscillation.  This suppression is determined by a numerical fit, parameterised by a Gaussian parameter $\sigma$ which describes the mass distribution of the black holes or Q-balls.  Their analysis uses piecewise expressions for the scale factor, Hubble parameter, and Green's function.  In contrast, in our analysis we have replaced all three with smoothly varying expressions.  Because of this, no further normalisation factor is needed in $\Phi$; its decline can be observed in the gradual line in figure~\ref{fig:phi}.  Because we consider a tanh decay profile to interpolate between fast and slow transitions, as opposed to a Gaussian distribution of decaying matter, it is difficult to compare our results directly; there is no direct equivalence between $\beta$ and $\sigma$ values.  We note, however, that figure~6 of ref.~\cite{Inomata:pbh} shows a resonance peak for all $\sigma$ values, whereas we clearly see the resonance disappear as the transition time increases.  This is to be expected as they explicitly note that they only consider mass distributions sufficiently small that the MD and cross terms in eq.~\eqref{eq:split} are negligible.  Therefore, this is the first work to explore the intermediate regime between the rapid and gradual transitions.

\section{Concluding remarks}
\label{sec:conclusions}

The GW signal from an early MD epoch has attracted a great deal of interest, particularly as new GW detectors are designed.  Detailed calculations of finite-duration transitions between matter and radiation domination are necessary, particularly because a fast transition shows a resonant peak while a slow transition displays a very precise cancellation.  In this work we have calculated the GW signal carefully for these intermediate transitions, avoiding all step-function approximations, whether in the scale factor or Green's function.  Our results show that the GW signal interpolates between the instantaneous limit and the gradual limit quickly enough that there is reason to be optimistic about the prospects of distinguishing between different sources of matter domination. Ultimately, one might be able to reconstruct three features of the scenario responsible for an equation of state: when the period of matter domination began, how long it lasted and how quickly it transitioned to radiation. 

In this work, we have used eq.~\eqref{eq:decay} to interpolate between the fast and slow transition regimes.  Of course, physical scenarios, such as those with primordial black holes or Q-balls, will not have a tanh profile.  We leave extending this analysis to these physically motivated scenarios for future work, as the calculation of the equivalent of eq.~\eqref{eq:decay} is complicated due to gauge effects between the synchronous and Newtonian gauges. 

\vspace{2mm}
MP was supported by an Australian Government Research Training Program (RTP) scholarship and a Monash Graduate Excellence scholarship (MGES).  CB acknowledges support from the Australian Research Council via the Discovery project DP210101636.

\bibliography{references}

\appendix
\section{Evolution of Perturbations in the Newtonian Gauge}
\label{app:perturbations}
Here we derive the equations of motion for perturbations in the Newtonian gauge for a matter component decaying to radiation with a generic time dependent decay rate.  The decay rate takes the simple form given in equation \eqref{eq:decay} in the synchronous gauge, comoving with the matter.  The perturbed metric in the synchronous gauge, including second order tensor perturbations, is given by
\begin{equation}
    ds^2 = a^2\left[-d\eta^2+\left(\delta_{ij}+H_{ij}+\frac{h_{ij}}{2}\right)dx^idx^j\right]
\end{equation}
where the scalar part of the metric is defined in Fourier space as
\begin{equation}
    H_{ij}=\hat{k}_i\hat{k}_j\gamma+\left(\hat{k}_i\hat{k}_j-\frac{1}{3}\delta_{ij}\right)6\epsilon.
\end{equation}
Here $\hat{\boldsymbol{k}}=\boldsymbol{k}/k$ and we have adopted the labelling from references~\cite{Inomata:gradual,Inomata:pbh}, where the fields $\gamma$, $\epsilon$, $\Phi$ and $\Psi$ correspond to $h$, $\eta$, $\psi$ and $\phi$ respectively from references~\cite{Ma:1995ey,Poulin:2016nat}.  While we have included the second order tensor perturbations $h_{ij}$, they do not contribute to the following discussion as it is limited to first order.

The perturbations in the Newtonian and synchronous gauges are related by
\begin{align}
\label{eq:GTphi}
    \Phi&=\mathcal{H}\alpha+\alpha' \\
\label{eq:GTpsi}
    \Psi&=\epsilon-\mathcal{H}\alpha\\
\label{eq:GTdelta}
    \delta^{(N)}_{m/r}&=\delta^{(S)}_{m/r}+\frac{\rho_{m/r}'}{\rho_{m/r}}\alpha\\
\label{eq:GTtheta}
    \theta_{m/r}^{(N)}&=\theta_{m/r}^{(S)}+k^2\alpha
\end{align}
where $\alpha=(6\epsilon+\gamma)'/(2k^2)$ and going forward we will suppress superscripts for the synchronous gauge.  Following the approach of references~\cite{Poulin:2016nat,Ma:1995ey}, the evolution of the phase space distribution for the decaying matter component $f_m(\eta,\boldsymbol{x},q,\hat{\boldsymbol{n}})$ can be described by the Boltzmann equation
\begin{equation}
    \frac{df_m}{d\eta}=\frac{\partial f_m}{\partial\eta}+\frac{\partial f_m}{\partial x^i}\frac{dx^i}{d\eta}+\frac{\partial f_m}{\partial q}\frac{dq}{d\eta}+\frac{\partial f_m}{\partial n^i}\frac{dn^i}{d\eta}=-a\Gamma(\eta)f_m
\end{equation}
where $\boldsymbol{q}=a\boldsymbol{p}=q\hat{\boldsymbol{n}}$ is the comoving 3-momentum of the fluid.  We can split the distribution into a background and first-order perturbation part
\begin{equation}
    f_m(\eta,\boldsymbol{x},q,\hat{\boldsymbol{n}})=\bar{f}_m(\eta,q)\left(1+\psi_m(\eta,\boldsymbol{x},q,\hat{\boldsymbol{n}})\right)  .
\end{equation}
The components of the stress energy tensor can then be obtained as phase space integrals and moments of $f_m$
\begin{align}
    T^0{}_{\!0}&=-\rho_m(1+\delta_m)=-a^{-4}\int dqd\Omega\; q^2\sqrt{q^2+m^2a^2}\bar{f}_m(1+\psi_m)\\
    \label{eq:moment}
    T^0{}_{\!i}&=\rho_m v_{mi}=a^{-4}\int dqd\Omega\; q^2qn_i\bar{f}_m\psi_m\\
    T^i{}_{\!j}&=0=a^{-4}\int dqd\Omega\; q^2\frac{q^2n_i n_j}{\sqrt{q^2+m^2a^2}}\bar{f}_m(1+\psi_m).
\end{align}
where $v_{mi}$ is the velocity perturbation of the matter, with $\theta_m=ik^i v_{mi}$ above.

At zeroth order, after integrating over the phase space, the Boltzmann equation reduces to the continuity equation eq.~\eqref{eq:fried_m}.  At first order, after transforming to Fourier space and switching to the physical momentum $\boldsymbol{p}=p\hat{\boldsymbol{n}}$, the Boltzmann equation becomes
\begin{equation}
\label{eq:Boltzmannsimp}
    \frac{\partial (\bar{f}_m\psi_m)}{\partial\eta}+i\frac{p}{E}(\boldsymbol{k}\cdot\hat{\boldsymbol{n}})\bar{f}_m\psi_m+p\frac{\partial \bar{f}_m}{\partial p}\left(\epsilon'-\frac{\gamma'+6\epsilon'}{2}(\hat{\boldsymbol{k}}\cdot\hat{\boldsymbol{n}})^2\right)-\mathcal{H}p\frac{\partial (\bar{f}_m\psi_m)}{\partial p}=-a\Gamma(\eta)\bar{f}_m\psi_m
\end{equation}
where $E=\sqrt{p^2+m^2}$.  Integrating over phase space and applying the background continuity equation eq.~\eqref{eq:fried_m} results in
\begin{equation}
\label{eq:deltamsync}
    \delta_m'=-\frac{\gamma'}{2}-\theta_m.
\end{equation}
If we instead take the first moment and then divergence of eq.~\eqref{eq:Boltzmannsimp} (i.e. apply eq.~\eqref{eq:moment} then we get the equation for the velocity divergence
\begin{equation}
\label{eq:thetamsync}
    \theta_m'=-\mathcal{H}\theta_m,
\end{equation}
which, provided the matter begins at rest, ensures $\theta_m^{(S)}=0$ as expected.  Lastly from energy and momentum conservation $\nabla_{\mu}T^{\mu\nu}=0$, the corresponding equations for the radiation component are
\begin{align}
\label{eq:deltarsync}
    \delta_r'=-\frac{4}{3}\left(\theta_r+\frac{h'}{2}\right)+a\Gamma(\eta)\frac{\rho_m}{\rho_r}(\delta_m-\delta_r)\\
\label{eq:thetarsync}
    \theta_r'=\frac{k^2}{4}\delta_r-a\Gamma(\eta)\frac{3\rho_m}{4\rho_r}\left(\frac{4}{3}\theta_r-\theta_m\right).
\end{align}

We can now turn to the issue of transforming these equations to the Newtonian gauge.  We note that ratio $\rho_{m/r}'/\rho_{m/r}$ can be replaced using the continuity equations eqs.~\eqref{eq:fried_m} and \eqref{eq:fried_r} which introduces a time derivative of the decay rate when transforming $\delta_{m/r}'$.  Applying the gauge transformation rules eqs.~\eqref{eq:GTphi} to \eqref{eq:GTtheta} to the equations of motion for the perturbations in the synchronous gauge eqs.~\eqref{eq:deltamsync} to \eqref{eq:thetarsync} results in
\begin{align}
    {\delta^{(N)}_m}'&=-\theta^{(N)}_m+3\Phi'-a\Gamma(\eta)\Psi-a\alpha\Gamma'(\eta) \\
    {\theta^{(N)}_m}'&=-\mathcal{H}\theta^{(N)}_m+k^2\Psi\\
    {\delta^{(N)}}'&=-\frac{4}{3}\left(\theta^{(N)}_r-3\Phi'\right)+a\Gamma(\eta)\frac{\rho_m}{\rho_r}\left(\delta^{(N)}_m-\delta^{(N)}_r+\Psi\right)+a\Gamma'(\eta)\frac{\rho_m}{\rho_r}\alpha\\
    {\theta^{(N)}_r}'&=\frac{k^2}{4}\delta^{(N)}_r+k^2\Psi-a\Gamma(\eta)\frac{3\rho_m}{4\rho_r}\left(\frac{4}{3}\theta^{(N)}_r-\theta^{(N)}_m\right)  .
\end{align}
The remaining factors of $\alpha$ can be expressed in terms of Newtonian gauge quantities by observing that in the synchronous gauge where $\theta^{(S)}_m=0$ eq.~\eqref{eq:GTtheta} implies
\begin{equation}
    \alpha=\frac{\theta^{(N)}_m}{k^2},
\end{equation}
which, after setting $\Psi=\Phi$ for negligible anisoptropic stress, results in the equations of motion we use in eqs.~\eqref{eq:deltam} to \eqref{eq:thetar}.  Lastly we note that setting $\Gamma(\eta)\equiv\Gamma$ does in fact result in the expressions of references~\cite{Poulin:2016nat,Inomata:gradual}.

\section{Technical Details of Code}
\label{app:code}
There are several systems of differential equations that need to be solved numerically in the process of calculating the GW spectra.  These are the Friedmann equations, eqs.~\eqref{eq:fried_m} to \eqref{eq:fried_a}, the GW Green's function, eq.~\eqref{eq:greens}, and the evolution of the perturbations, eqs.~\eqref{eq:deltam} to \eqref{eq:phiprime}.  For this purpose we use \texttt{Odeint} from the \texttt{Boost} \texttt{C++} library.  Specifically we use the \texttt{runge\_kutta\_dopri5} solver with an absolute tolerance of $10^{-200}$ and a relative tolerance of $10^{-11}$.  The absolute tolerance of $10^{-200}$ is necessary to prevent the solver stalling once $\rho_m$ decays to values close to the smallest representable machine number.

To perform integrals numerically we used Gauss-Legendre quadratures which we generate using the algorithm described in ref.~\cite{Bogaert:quads}.  For the integral over $\bar{\eta}$ in eq.~\eqref{eq:I} we note that once $\Phi$ begins to oscillate it does so with period $2\pi\sqrt{3}/k$ and we ensure that the integrand is sampled densely enough that there are approximately 50 points per period.

For the integral over $u$ and $v$ in eq.~\eqref{eq:spectra} we first transform to the variables $s=u-v$ and $t=u+v-1$ which gives~\cite{Kohri:2018awv}
\begin{equation}
    \overline{\mathcal{P}_h(\eta,k)}=2\int_{0}^{\infty}dt\int_{-1}^{1}ds\left(\frac{t(2+t)(s^2-1)}{(1-s+t)(1+s+t)}\right)^2\overline{I^2(u,v,k,\eta)}\mathcal{P}_{\zeta}(uk)\mathcal{P}_{\zeta}(vk)
\end{equation}
where the remaining factors of $u$ and $v$ can be expressed in terms of $s$ and $t$ as $u=(t+s+1)/2$ and $v=(t-s+1)/2$.  This form has several benefits.  Firstly, the limits of the inner integral over $s$ are no longer dependent on the outer integration variable.  Additionally the integrand is symmetric in $s$ which we will exploit to only have to perform the integral from 0 to 1.  Lastly, $I$ has a pole at $t=\sqrt{3}-1$, which is responsible for the resonant peak in the rapid transition spectra.

We found that the integrand was only very mildly dependent on $s$ and so we could perform the integral over $s$ from 0 to 1 accurately with only 5 quadrature points.  Increasing the number of quadrature points in $s$ to 10 only impacted the result at less than the $\mathcal{O}(0.01\%)$ level in both the rapid and gradual cases.  For the integral over $t$ we divide the integration region into $0\leq t<\sqrt{3}-1$, $\sqrt{3}-1<t\leq 2(\sqrt{3}-1)$ and $t>2(\sqrt{3}-1)$.  To ensure we account for the important contribution from the pole we sample the regions above and below the pole with 40 quadrature points each.  The remainder of the integral was computed with 20 more quadrature points.  We found that this distribution of quadratures could have as much as 5 percent error for some values of $k$ for a rapid transition when compared to using 200 points either side of the pole and 100 in the remainder.  The error in the gradual case was negligible due to $I$ becoming approximately independent of $s$ and $t$.  Using this larger number of quadrature points would significantly increase the run-time and a 5 percent error is adequate to compare with the gradual and rapid limits from previous work and make qualitative statements about the spectra and potential detectability.

\end{document}